\documentclass[aip,twocolumn,jcp]{revtex4}

\usepackage{graphicx}              
\usepackage{amssymb}
\usepackage{amsmath}
\usepackage{braket}
\usepackage{mathrsfs}

\newcommand{\Tr}[1]{\mathrm{Tr\left[#1\right]}}
\newenvironment{acknowledgment}{{\flushleft \bf Acknowledgments:}}{}
\newcommand{\R}{\mathbf{r}}

\begin{document}

\title{Frozen density embedding with non-integer subsystems' particle numbers }

\author{Eduardo Fabiano}\email{eduardo.fabiano@nano.cnr.it}
\affiliation{National Nanotechnology Laboratory (NNL), Istituto
Nanoscienze-CNR, Via per Arnesano 16, I-73100 Lecce, Italy }
\author{Savio Laricchia}
\affiliation{National Nanotechnology Laboratory (NNL), Istituto
Nanoscienze-CNR, Via per Arnesano 16, I-73100 Lecce, Italy }
\author{Fabio Della Sala}
\affiliation{National Nanotechnology Laboratory (NNL), Istituto 
Nanoscienze-CNR,
Via per Arnesano 16, I-73100 Lecce, Italy }
\affiliation{Center for Biomolecular Nanotechnologies @UNILE, Istituto 
Italiano di Tecnologia,
Via Barsanti, I-73010 Arnesano, Italy}
\date{\today}

\begin{abstract}
We extend the frozen density embedding theory to non-integer subsystems'
particles numbers. Different features of this formulation are discussed,
with special concern for approximate embedding calculations. In particular,
we highlight the relation between the non-integer particle-number
partition scheme and the resulting embedding errors.
Finally, we provide a discussion of the implications of the
present theory for the derivative discontinuity issue and the
calculation of chemical reactivity descriptors. 
\end{abstract}

\maketitle

\section{Introduction}
Subsystem approaches \cite{senatore86,cortona91,yang91,cortona94,yang95,sugiki03,fedorov04,huang06,weso_chap06,cohen07,cohen07_2,tomoko07,wang08,zhao08,he09,elliott10,huang11,huang11_2,gordon12,severo12,mamby12,orbital_free_book}
are powerful tools in the context of density functional theory \cite{hk,parr_book}.
In particular, the frozen density embedding (FDE) theory  and its related
methods \cite{senatore86,cortona91,weso_chap06,wesolowski93,neugebauer05,jacob06,jacob08,wesolowski08,neugebauer09_2,hyb_emb,kaminski10,fux10,wesolowski11,fux11,pavanello11,lhf_emb,aquilante11,silva12,emb_energy,jacobrev}, 
emerged in the last years as simple and powerful approaches to treat, in
an in principle exact manner, non-covalent complexes 
\cite{jacob06,pavanello11,neugebauer05_2,jacob05,kevorkyants06,dulak07,dulak07_2,neugebauer09,gotz09,kaminski10,apbe,apbek,fradelos11,fradelos11_2,emb_energy,emb_ct}.

Within the FDE method the electron density is partitioned into several  subsystem's contributions
that are constrained to sum up to the total electron density. The ground-state density is then
obtained by minimizing the total energy of the system with respect to one of the subsystem's densities
keeping the other ones fixed (i.e., frozen). In this way, in principle, the full solution of the
Hohenberg-Kohn variational problem \cite{hk} is obtained. If auxiliary orbitals are introduced to expand
the electron density, the FDE problem can be recast into the familiar framework of the Kohn-Sham (KS)
method \cite{ks}. In this case the ground-state solution is given by a set of coupled
KS equations with constrained electron density (KSCED) \cite{wesolowski93}.

The partition of the density, which is at the base of the FDE procedure, is in principle
arbitrary and any convenient choice can be made for the system under examination.
However, the actual formalism of the FDE theory, both in the orbital-free and the
orbital-based formulation, is founded on a pure-state representation of the electron
densities. Therefore, 
only subsystem densities with integer number of electrons are allowed.  
The integer occupation of subsystems limits
the choice of possible frozen densities and, in addition, has important
consequences from a more formal point of view. In fact, non-integer
particle numbers have been shown to be of extreme importance for the
deep comprehension and the development of density functional theory
\cite{perdew82,janak78,yang12,cohen12,cohen08,cohen08_2,frac_scaling}. 
In particular, we mention the fact that the use
of non-integer numbers of particles is intrinsically related to the
fundamental issue of the derivative discontinuity of density
functional theory \cite{perdew82,cohen12} and stands at 
the basis of all definitions in chemical reactivity theory 
\cite{cohen07,parr_book,crt_book,chermette99,geerlings03}.
In addition, concerning this latter aspect, Ref. \cite{cohen07}
showed the importance of both non-integer occupations and environmental
effects for a proper definition of chemical descriptors.
Thus, in order to be able to have deep insight into the
fundamental properties of the embedding potential or
access chemical reactivity descriptors
for the embedded subsystems via the FDE approach, it is necessary to 
have a proper FDE theory for non-integer particle numbers.

We acknowledge that other embedding theories exist where non-integer
particle numbers are allowed \cite{elliott10,huang11,huang11_2,elliott09,tang12}.
Namely, the partition density-functional theory (PDFT) \cite{elliott10,elliott09} and the
potential-functional embedding theory of Huang and Carter \cite{huang11,huang11_2}. 
The latter is not a density-based theory, since the total electronic energy is formulated 
as a functional of the embedding potential rather than of the subsystems’ densities. 
Moreover, in both cases no frozen
density is considered but the solution is found by variational minimization of
all the subsystems' densities and particle numbers. Thus, the 
non-integer occupation of a given subsystem cannot be considered as a
parameter inherent with the chosen partition scheme, as in FDE, 
but it is automatically 
determined by the self-consistent solution of the embedding problem.
Furthermore, we note that the minimization of the total electronic energy 
with respect to the subsystems' occupations may arise some conceptual 
problems in practical calculations where an approximate density functional 
is employed for the kinetic energy, because the true and approximate energy 
functionals may be minimized by different sets of occupation numbers \cite{mosquera13}.

In this paper we 
focus on the FDE formalism and consider its extension to
non-integer subsystems' particle numbers, by considering a zero-temperature
grand-canonical ensemble formalism \cite{perdew82}. The implications
of the developed theory are then discussed to highlight the significance
of non-integer subsystems' particle numbers, especially
in approximate embedding calculations. Furthermore, we 
shortly comment on the relevance of the present approach for the 
comprehension of the role of the derivative discontinuity in the
FDE context as well as for the definition of well-posed 
chemical descriptors for embedded subsystems.

\section{Frozen density embedding with ensemble subsystem densities}
\label{sec:theory}
Within the Levy's constrained search \cite{levy79,lieb83} it is well
established that the ground-state energy of any $N$-electron system can
be expressed as
\begin{equation}\label{e1}
E_0 = \min_{\rho\rightarrow  N}\left\{\min_{\Psi\rightarrow\rho}
\braket{\Psi|\hat{T}+\hat{V}_{ee}|\Psi}+\int\rho(\R)v_{\text{ext}}(\R)d\R\right\}\ ,
\end{equation}
where $N$ is any integer number, 
$\rho$ is an $N-$representable electron density, $\Psi$ is an
$N$-particle 
wave function yielding the density $\rho$, $\hat{T}$ is the kinetic energy 
operator, $\hat{V}_{ee}$ is the electron-electron interaction operator, and
$v_{\text{ext}}$ is an external potential (usually the nuclear potential).
The functional
\begin{equation}\label{e2}
F[\rho] =  \min_{\Psi\rightarrow\rho}\braket{\Psi|\hat{T}+\hat{V}_{ee}|\Psi}
\end{equation}
is called the Hohenberg-Kohn universal functional and its existence is
guaranteed by the first theorem of Hohenberg and Kohn \cite{hk}.

Now we consider a partitioning of the density into $\rho=\rho_{A\omega}+\rho_{B\omega}$,
where $\rho_{A\omega}$ and $\rho_{B\omega}$ are the ensemble
$(N_A+\omega)$- and $(N_B-\omega)$-representable electron 
densities
\begin{equation}\label{e3}
\rho_{A\omega}  =  \Tr{\hat{\Gamma}_{A\omega}\hat{\rho}} \quad ; \quad \rho_{B\omega}  =  \Tr{\hat{\Gamma}_{B\omega}\hat{\rho}}\ .
\end{equation}
Here $N_A$ and $N_B$ are two integer numbers such that
$N_A+N_B=N$, $\omega\in[-1,1]$, $\hat{\rho}$ is the density operator
in second quantization representation, and 
\begin{equation}\label{egammaa}
\hat{\Gamma}_{A\omega} =  \pm\omega|\Psi_{A,N_A\pm 1}\rangle\langle\Psi_{A,N_A\pm 1}| + (1\mp\omega)|\Psi_{A,N_A}\rangle\langle\Psi_{A,N_A}| 
\end{equation}
\begin{equation}
\hat{\Gamma}_{B\omega} = \pm\omega|\Psi_{B,N_B\mp 1}\rangle\langle\Psi_{B,N_B\mp 1}|  +  (1\mp\omega)|\Psi_{B,N_B}\rangle\langle\Psi_{B,N_B}|
\end{equation}
are the density operators describing
the subsystems $A$ and $B$, respectively, with
$\Psi_{i,M}$ being an $M$-electron wave function for subsystem $i$
and the upper and lower signs in $\pm$ and $\mp$ corresponding
to $\omega>0$ and $\omega<0$ respectively. 
Furthermore, 
we assume that $\rho_{B\omega}$ is fixed to some
given trial value respecting minimal conditions for a physical 
subsystem density (see subsection \ref{sec_scf}). 

According to our partition scheme we define the following functionals
\begin{eqnarray}
\label{e5}
&&F[\rho_{A\omega}]  =  \min_{\hat{\Gamma}_{A\omega}\rightarrow\rho_{A\omega}}
\Tr{\hat{\Gamma}_{A\omega}\left(\hat{T}+\hat{V}_{ee}\right)} \\
\label{e6}
&&F[\rho_{B\omega}]  =  \min_{\hat{\Gamma}_{B\omega}\rightarrow\rho_{B\omega}}
\Tr{\hat{\Gamma}_{B\omega}\left(\hat{T}+\hat{V}_{ee}\right)} \\
\label{e7}
&&F^{nadd}[\rho_{A\omega},\rho_{B\omega}] = F[\rho_{A\omega}+\rho_{B\omega}] - F[\rho_{A\omega}] -F[\rho_{B\omega}]\ ,
\end{eqnarray}
where $F[\rho_{A\omega}+\rho_{B\omega}]=F[\rho]$ is independent from $\omega$ and
is given by Eq. (\ref{e2}).
The ground-state energy of Eq. (\ref{e1}) can be thus written
\begin{eqnarray}
\nonumber
E_0 & = & \min_{\rho_{A\omega}\rightarrow N_A+\omega}\biggl\{F[\rho_{A\omega}]+F[\rho_{B\omega}]+ F^{nadd}[\rho_{A\omega},\rho_{B\omega}] \nonumber \\
\label{e8}
 && +\int\left[\rho_{A\omega}(\R)+\rho_{B\omega}(\R)\right]v_{\text{ext}}(\R)d\R\biggr\}\ ,
\end{eqnarray}
where the minimization over $\rho$ was substituted by the minimization
over $\rho_{A\omega}$ in consideration of the fact that $\rho_{B\omega}$ is fixed.
Note that Eq. (\ref{e8}) is in fact fully equivalent to Eq. (\ref{e1}),
despite the fact that $\rho_{B\omega}$ and $\omega$ are arbitrarily fixed, 
because it always admits the
formal solution $\rho_{A\omega}^{min}=\rho^{min}-\rho_{B\omega}$, 
where $\rho^{min}$ is the density minimizing Eq. (\ref{e1})
\cite{weso_chap06}.

According to Eq. (\ref{e8}), we have to minimize the functional
\begin{eqnarray}
\nonumber
\mathscr{L}[\rho_{A\omega}] & = & F[\rho_{A\omega}]+F[\rho_{B\omega}]+F^{nadd}[\rho_{A\omega},\rho_{B\omega}] \nonumber \\ 
&+& \int\left[\rho_{A\omega}(\R)+\rho_{B\omega}(\R)\right]v_{\text{ext}}(\R)d\R \nonumber \\ 
&-& \label{e9}
 \mu\left(\int\Tr{\hat{\Gamma}_{A\omega}\hat{\rho}}d\R-N_A-\omega\right)\ , \nonumber \\
\end{eqnarray}
where $\mu$ is the Lagrange multiplier associated with the condition
$\rho_{A\omega}\rightarrow  N_A+\omega$. 
To minimize the functional $\mathscr{L}$
we consider a small variation of the density operator 
$\hat{\Gamma}_{A\omega}\rightarrow\hat{\Gamma}_{A\omega}+\delta\hat{\Gamma}_{A\omega}$ so that
$\rho_{A\omega}\rightarrow\rho_{A\omega}+\delta\rho_{A\omega}$ with 
$\delta\rho_{A\omega}=\Tr{\delta\hat{\Gamma}_{A\omega}\hat{\rho}}$.
Thus, retaining only the first-order terms, we obtain
\begin{equation}\label{e10}
\int\left(\frac{\delta F[\rho_{A\omega}]}{\delta\rho_{A\omega}(\R)}
+\frac{\delta F^{nadd}[\rho_{A\omega},\rho_{B\omega}]}
{\delta\rho_{A\omega}(\R)}+v_{\text{ext}}(\R)-\mu\right)\delta\rho_{A\omega}(\R)d\R = 0\ ,
\end{equation}
which yields the Euler equation
\begin{equation}\label{e11}
\frac{\delta F[\rho_{A\omega}]}{\delta\rho_{A\omega}(\R)}+\frac{\delta F^{nadd}[\rho_{A\omega},\rho_{B\omega}]}{\delta\rho_{A\omega}(\R)}+v_{ext}(\R) = \mu\ .
\end{equation}
If we partition the external potential as $v_{\text{ext}}=v_{\text{ext}}^A+
v_{\text{ext}}^B$, we can finally write
\begin{equation}\label{e12}
\frac{\delta F[\rho_{A\omega}]}{\delta\rho_{A\omega}(\R)}+v_{\text{ext}}^A(\R)+
v_{emb}^A[\rho_{A\omega};\rho_{B\omega}](\R) = \mu\ ,
\end{equation}
where we defined
\begin{equation}\label{e13}
v_{emb}^A[\rho_{A\omega};\rho_{B\omega}](\R) \equiv \frac{\delta F^{nadd}[\rho_{A\omega},\rho_{B\omega}]}{\delta\rho_{A\omega}(\R)}+
v_{\text{ext}}^B(\R)\ .
\end{equation}
The embedding potential $v_{emb}^A$ is therefore the external potential 
needed to constrain the solution of the subsystem $A$ to yield the
electron density of the total system through the condition
$\rho=\rho_{A\omega}+\rho_{B\omega}$.

\subsection{Kohn-Sham formulation}
\label{ks_sec}
To develop a Kohn-Sham formulation for our problem
we introduce, for each value of $\omega$, a
reference ensemble non-interacting system with
external potential $w$ such that the 
ensemble non-interacting and interacting densities
are the same. According to Eq. (\ref{egammaa}), such
a system is described by the density operator
\begin{equation}\label{e19}
\hat{\Gamma}_{A\omega}^{KS} = \pm\omega|\Phi^{(\omega)}_{A,N_A\pm 1}\rangle\langle\Phi^{(\omega)}_{A,N_A\pm 1}| + (1\mp\omega)|\Phi^{(\omega)}_{A,N_A}\rangle\langle\Phi^{(\omega)}_{A,N_A}|\ ,
\end{equation}
where $\langle\R_1\ldots\R_M|\Phi^{(\omega)}_{A,M}\rangle$ is an $M$-particle 
Slater determinant, depending implicitly on $\omega$
(for each value of $\omega$ a different external potential
$w$ will be needed to guarantee the equality of the ensemble 
non-interacting and interacting densities; throughout the paper 
the superscript $^{(\omega)}$ will be used to denote this implicit dependence).
Thus, the reference system has a non-interacting kinetic energy
\begin{equation}\label{e14}
T_s[\rho_{A\omega}] = \min_{\hat{\Gamma}^{KS}_{A\omega}\rightarrow\rho_{A\omega}}\Tr{\hat{\Gamma}^{KS}_{A\omega}\hat{T}}
\end{equation}
and its ground state is described by the Euler equation
\begin{equation}\label{e17}
\frac{\delta T_s[\rho_{A\omega}]}{\delta\rho_{A\omega}(\R)} + w(\R) = \mu\ .
\end{equation}
Comparison of Eqs. (\ref{e12}) and (\ref{e17}) allows to identify 
$w(\R)=v_s^A(\R)+v_{emb}^A(\R)$ where we defined the KS potential
\begin{equation}\label{e18}
v_s^A(\R) = \int\frac{\rho_{A\omega}(\R_1)}{|\R-\R_1|}d\R_1 + 
\frac{\delta E_{xc}[\rho_{A\omega}]}{\delta\rho_{A\omega}(\R)} + v_{\text{ext}}^A(\R) \ ,
\end{equation}
with $E_{xc}[\rho_{A\omega}] = F[\rho_{A\omega}] - T_s[\rho_{A\omega}] - J[\rho_{A\omega}]$ 
and $J$ the classical Coulomb energy.

Practical equations for ground-state calculations are obtained 
considering that the KS Slater determinants entering in Eq. 
(\ref{e19}) are constructed with
orbitals that are solutions of the Schr\"odinger equation
for a single particle in an external potential $w$. Thus,
the ground-state for subsystem $A$ is given by the KSCED
\begin{equation}\label{e21}
\left(-\frac{1}{2}\nabla^2+v^A_{s}[\rho_{A\omega}]+
v_{emb}^A[\rho_{A\omega};\rho_{B\omega}]\right)\phi^{(\omega)}_{A,i} = \epsilon^{(\omega)}_{A,i}\phi^{(\omega)}_{A,i}\ ,
\end{equation}
with the electron density computed as
\begin{equation}\label{mmm1}
\rho_{A\omega}=\Tr{\hat{\Gamma}^{KS}_{A\omega}\hat{\rho}}=\sum_if_i|\phi_{A,i}^{(\omega)}|^2 \ ,
\end{equation}
where the occupation numbers are
\begin{eqnarray}\label{mmm2}
f_i & = & \left\{
\begin{array}{lc}
1 & i\leq N_A\\
\omega & i = N_A+1\\
0 & i > N_A+1\\
\end{array}\right. \ \ \mathrm{for}\ \omega>0 \quad ; \\
f_i & = & \left\{
\begin{array}{lc}
1 & i\leq N_A-1\\
1+\omega & i = N_A\\
0 & i > N_A\\
\end{array}\right. \ \ \mathrm{for}\ \omega<0\ .
\end{eqnarray}

In the KS scheme introduced above the embedding potential
of Eq. (\ref{e13}) can be recast into a more practical form by writing
\begin{eqnarray}\label{e22}
v_{emb}^A[\rho_{A\omega};\rho_{B\omega}](\R) & = & v_{\text{ext}}^B(\R)
 + \int\frac{\rho_{B\omega}(\R_1)}{|\R-\R_1|}d\R_1 
 \nonumber \\ &+& \frac{\delta T^{nadd}_s[\rho_{A\omega};\rho_{B\omega}]}{\delta\rho_{A\omega}(\R)} 
+ \frac{\delta E^{nadd}_{xc}[\rho_{A\omega};\rho_{B\omega}]}{\delta\rho_{A\omega}(\R)} \ , \nonumber \\
\end{eqnarray}
where
\begin{eqnarray}
&&T^{nadd}_s[\rho_{A\omega};\rho_{B\omega}] = T_s[\rho_{A\omega}+\rho_{B\omega}]-T_s[\rho_{A\omega}]-T_s[\rho_{B\omega}]
\nonumber \\
&&E^{nadd}_{xc}[\rho_{A\omega};\rho_{B\omega}] = E_{xc}[\rho_{A\omega}+\rho_{B\omega}]-E_{xc}[\rho_{A\omega}]- E_{xc}[\rho_{B\omega}] \nonumber \\
\end{eqnarray}
are the non-additive non-interacting kinetic and exchange-correlation
energy functionals, respectively.
Appropriate approximations must be used to express $T_s$ and $E_{xc}$
for practical calculations (see section \ref{sec:approx_emb}).
In particular we note that, while the functionals 
$T_s[\rho_{A\omega}+\rho_{B\omega}]=T_s[\rho]$ and
$E_{xc}[\rho_{A\omega}+\rho_{B\omega}]=E_{xc}[\rho]$ 
are defined for pure-state densities, so that
conventional approximations \cite{scuseria06,wesolowski97,gotz09,apbek}
can be employed straightforwardly,
the functionals $T_s[\rho_{A\omega}]$ and $E_{xc}[\rho_{A\omega}]$ 
are defined for ensemble densities, 
thus special care may be needed in their approximation
(see e.g. Refs. \onlinecite{ullrich01,kraisler13}).

\subsection{Full self-consistent solution}
\label{sec_scf}
In the previous sections we considered $\rho_{B\omega}$ fixed to any arbitrary
function satisfying the minimal requirements for a physical
subsystem density \cite{parr_book}, i.e. everywhere we must have
$0\leq \rho_{B\omega}\leq\rho$, $0<\int\rho_{B\omega}\, d\R<N$ and also 
$\int |\nabla\rho_{B\omega}^{1/2}|^2\, d\R < \infty$.
In this way only $\rho_{A\omega}$ was variationally optimized to yield 
the full ground-state density. This condition, although it may 
result advantageous in some situations, leads to the problem 
of the choice of an optimal $\rho_{B\omega}$.
In fact, even simple choices of $\rho_{B\omega}$ turn out to be problematic.
For example, the use of the electron density of the isolated subsystem $B$
generally violates the condition $\rho_{B\omega}\leq\rho$ 
in some regions of space \cite{kiewisch08}. 
Moreover, if the Kohn-Sham scheme is used for solution, as it is often the 
case, the frozen density must satisfy the additional condition that
$\rho_{A\omega}=\rho-\rho_{B\omega}$ 
is a $v_s$-representable electron density, that means
that it must be the ground-state associated with a KS 
potential. This condition is of course very hard to fulfill a priori,
because the total density $\rho$ is not known in embedding calculations.

A failure of the frozen $\rho_{B\omega}$ to fulfill all the requirements
described above, will prevent the embedding calculation to
converge to the correct solution.
Therefore, it may be preferable instead to determine both 
$\rho_{A\omega}$ and $\rho_{B\omega}$
through a variational procedure.

To this end we can reconsider Eq. (\ref{e8}) and write it as
\begin{eqnarray}\label{eq:totE}
\nonumber
E_0 & = & \min_{\substack{\rho_{A\omega}\rightarrow  N_A + \omega \\ \rho_{B\omega}\rightarrow  N_B-\omega}}
\left\{F[\rho_{A\omega}]+F[\rho_{B\omega}]+F^{nadd}[\rho_{A\omega},\rho_{B\omega}]+ \right. \\
\label{e23}
 &&\left. +\int\left[\rho_{A\omega}(\R)+\rho_{B\omega}(\R)\right]v_{\text{ext}}(\R)d\R\right\}\ ,
\end{eqnarray}
where now the minimization involves both densities and must be constrained
by the condition $N_A+N_B=N$. Thus, we have to minimize the functional
\begin{eqnarray}
\nonumber
&&\mathscr{L}[\rho_{A\omega},\rho_{B\omega}] = F[\rho_{A\omega}]+F[\rho_{B\omega}]+F^{nadd}[\rho_{A\omega},\rho_{B\omega}]+\\
\nonumber
&&\quad + \int\left[\rho_{A\omega}(\R)+\rho_{B\omega}(\R)\right]v_{\text{ext}}(\R)d\R -\\
\label{e24}
&&\quad - \mu\left(\int\Tr{\hat{\Gamma}_{A\omega}\hat{\rho}}d\R + 
\int\Tr{\hat{\Gamma}_{B\omega}\hat{\rho}}d\R -N\right)\ ,
\end{eqnarray}
Following the steps bringing from Eq. (\ref{e9}) to Eq. (\ref{e12}) we
finally find the set of coupled equations
\begin{eqnarray}
\label{e25}
\frac{\delta F[\rho_{A\omega}]}{\delta\rho_{A\omega}(\R)}+v_{\text{ext}}^A(\R) +
v_{emb}^A[\rho_{A\omega};\rho_{B\omega}](\R) & = & \mu \\
\label{e26}
\frac{\delta F[\rho_{B\omega}]}{\delta\rho_{B\omega}(\R)}+v_{\text{ext}}^B(\R) + 
v_{emb}^B[\rho_{B\omega};\rho_{A\omega}](\R) & = & \mu \ .
\end{eqnarray}
These equations, or the analogous set of KSCED, can be solved in
a freeze-and-thaw scheme \cite{wesolowski96} to yield a fully 
variational solution of the embedding problem.

Note however, that even though the densities are obtained through the
solution of the coupled equations (\ref{e25}) and (\ref{e26}), we are
still left with some arbitrarity in the choice of the subsystem densities.
In fact, in Eq. (\ref{e24}) there is only a constraint on the
total number of electrons, but no restrictions on the individual
values of $N_A+\omega$ and $N_B-\omega$. Therefore, for any system there exists a  
family of fully variational exact embedding solutions $\rho_{A\omega}$ and $\rho_{B\omega}$
which provide exactly the same total ground-state energies and densities.
Note that within the FDE theory, only the ground-state energy and density 
of the total system are meaningful, while the subsystems densities and energies are, 
in principle, not well (and uniquely) defined, unlike in PDFT\cite{elliott10}.

\subsection{Comments on approximate frozen density embedding}
\label{sec:approx_emb}
In practical applications of the Kohn-Sham method the
exchange-correlation (XC) energy (and the corresponding potential) need
to be approximated, because the exact form of this term is not known.
In the present paper and in most FDE applications only local/semilocal XC
approximations are considered. These suffice in fact for various applications
\cite{mukappa,goerigk10}. Moreover, they display an explicit dependence from the
electron density. With this choice, the XC part of the embedding potential
can be computed on the same level as the XC potential of the full KS
calculation and no additional approximation is needed in
the embedding calculations (for our purposes we can assume the XC 
approximation is used from the beginning in the construction of
the Hohenberg-Kohn universal functional of Eq. (\ref{e2})).

On the other hand, the non-interacting kinetic energy is treated exactly
within the KS scheme, but it needs to be approximated in the embedding
potential, since in this latter case  we cannot make use of auxiliary 
orbitals for the total embedded system, 
but rather require an explicitly density-dependent functional
(we remark that an alternative strategy based on the explicit 
construction of the kinetic potential using the numerical inversion procedures 
also exists \cite{fux10,silva12}, but is not applicable in practical calculations) .
Thus, local or semilocal approximations are introduced for the
non-additive kinetic energy in any practical calculation based
on the frozen density embedding method.
This approximation is the only source of errors in semilocal
embedding calculations, when compared with the corresponding 
KS results for the whole
system.

The inclusion of approximations in the embedding procedure has important 
consequences on the outcome of the embedding calculations
\cite{gritsenko_chap}. The first,
of course, is that the true ground-state energy and density cannot be recovered 
any more because Eqs. (\ref{e1}) and (\ref{e8}) (or (\ref{e1}) and (\ref{e23}))
minimize different functionals. More importantly, the final
result of the embedding calculation is no more independent on the
choice of the ``frozen'' density $\rho_{B\omega}$. In fact, while for the
exact kinetic energy the functional in Eq. (\ref{e8}) is
truly dependent only on the sum $\rho_{A\omega}+\rho_{B\omega}$, so that it
will yield always the same result whenever the sum is left unchanged,
for approximate kinetic energy functionals the functional of 
Eq. (\ref{e8}) really depends on $\rho_{A\omega}$ and $\rho_{B\omega}$
separately and will likely provide different results for different
choices of $\rho_{B\omega}$. 
This problem can be naturally mitigated by employing a full self-consistent
solution for the embedding densities, so that both $\rho_{A\omega}$ 
and $\rho_{B\omega}$ are
provided as a solution of the coupled equations (\ref{e25}) and (\ref{e26}).
However, as mentioned before, also in this case different solutions 
will be in general obtained for different choices of the charge 
partitioning scheme,
i.e. for different values of $\omega$. The investigation of this
dependence will be the main subject of the following sections.

\section{Effect of the non-integer particle numbers on the embedding density}
Consider an ensemble $v_s$-representable frozen density 
$\rho_{B\omega}$ which integrates to the fractional occupation $N_B-\omega$
with $\omega\geq 0$ (the case $\omega<0$ is analogous and will not be explicitly
considered in the following, except for the fact that some final formulas will be
stated for both cases when needed).
We assume that such a density can be written (which is valid also for a full
self-consistent calculation)
\begin{equation}\label{e28}
\rho_{B\omega} = \omega\rho_{B,N_B-1}^{(\omega)} + (1-\omega)\rho_{B,N_B}^{(\omega)} \ ,
\end{equation}
with the pure-state ($N_B-1$)- and $N_B$-electron densities defined as
\begin{equation}\label{eq:pureB}
\rho_{B,N_B-1}^{(\omega)} = \sum_i^{N_B-1}\left|\phi_{B,i}^{(\omega)}\right|^2\quad ; \quad \rho_{B,N_B}^{(\omega)} = \sum_i^{N_B}\left|\phi_{B,i}^{(\omega)}\right|^2\ ,
\end{equation}
where the $\phi_{B,i}^{(\omega)}$ are some orbitals solution of a 
single-particle equation of the form of Eq. (\ref{e21}).
Of course, the frozen density can be also written in the
alternative form given by Eqs. (\ref{mmm1}) and (\ref{mmm2}).

For an exact embedding calculation, the complementary subsystem 
density $\rho_{A\omega}$, with fractional occupation
$N_A+\omega$ will be $\rho_{A\omega} = \rho^{KS}-\rho_{B\omega}$.
This density is also ensemble $v_s$-representable. In fact,
direct substitution of this density into Eqs. (\ref{e18}), (\ref{e22}),
and (\ref{e21}) yields the equation
\begin{equation}\label{eert}
\left[-\frac{1}{2}\nabla^2+\frac{\delta T_s[\rho^{KS}]}{\delta\rho_{A\omega}} - 
\frac{\delta T_s[\rho_{A\omega}]}{\delta\rho_{A\omega}} + 
v_s^{KS}\right]\phi_{A,i}^{(\omega)}
 = \epsilon_{A,i}^{(\omega)}\phi_{A,i}^{(\omega)}\ ,
\end{equation}
where $v_s^{KS}$ denotes the Kohn-Sham potential of the full system.
The ($N_A+\omega$)-electron
density constructed from the orbitals satisfying Eq. (\ref{eert}) 
is, by construction, the one
that minimizes Eq. (\ref{e17}) (or equivalently Eq. (\ref{e12})).
Thus, it is by definition $\rho_{A\omega} = \rho^{KS}-\rho_{B\omega}$.

As described in Section \ref{ks_sec}, the electron density of subsystem $A$ is 
given by Eqs. (\ref{mmm1}) and (\ref{mmm2}).
Alternatively, we can write it as
\begin{equation}\label{e27}
\rho_{A\omega} = \omega\rho_{A,N_A+1}^{(\omega)} + (1-\omega)\rho_{A,N_A}^{(\omega)} \ ,
\end{equation}
with the pure-state densities defined as
\begin{equation}\label{qqq}
\rho_{A,N_A+1}^{(\omega)} = \sum_i^{N_A+1}\left|\phi_{A,i}^{(\omega)}\right|^2\quad ; \quad \rho_{A,N_A}^{(\omega)} = \sum_i^{N_A}\left|\phi_{A,i}^{(\omega)}\right|^2\ .
\end{equation}

Similar results hold also in the case of a full self-consistent FDE calculation,
where both $\rho_{A\omega}$ and $\rho_{B\omega}$ are optimized variationally.

Using the definitions (\ref{e28}) and (\ref{e27}) we can readily write for the
total embedding density
\begin{eqnarray}
\nonumber
\rho_{A\omega}+\rho_{B\omega} & = &\rho_{A,N_A}^{(\omega)}+\rho_{B,N_B}^{(\omega)} + \\
&& + \omega\left(\rho_{A,N_A+1}^{(\omega)}-\rho_{A,N_A}^{(\omega)}+
\rho_{B,N_B-1}^{(\omega)}-\rho_{B,N_B}^{(\omega)}\right)\ .
\nonumber \\
\end{eqnarray}
That is
\begin{equation}\label{www}
\rho_{A\omega}+\rho_{B\omega} =  \rho_{A,N_A}^{(\omega)}+\rho_{B,N_B}^{(\omega)}+ 
\omega\left(\left|\phi_{A,N_A+1}^{(\omega)}\right|^2 - 
\left|\phi_{B,N_B}^{(\omega)}\right|^2\right)\ .
\end{equation}
Expanding the orbitals and densities in powers of $\omega$, i.e.
\begin{eqnarray}
\phi_i^{(\omega)} & \approx & \phi_i^{(\omega=0)} + \omega\frac{\partial\phi_i^{(\omega)}}{\partial \omega}\Big|_{\omega=0^+} + 
\frac{\omega^2}{2}\frac{\partial^2\phi_i^{(\omega)}}{\partial \omega^2}\Big|_{\omega=0^+} + \cdots \\
\rho_i^{(\omega)} & \approx & \rho_i^{(\omega=0)} + \omega\frac{\partial\rho_i^{(\omega)}}{\partial \omega}\Big|_{\omega=0^+} + \frac{\omega^2}{2}\frac{\partial^2\rho_i^{(\omega)}}{\partial \omega^2}\Big|_{\omega=0^+} + \cdots
\ ,
\end{eqnarray}
it is then possible to express the total embedding density as a power series in $\omega$
\begin{equation}
\rho_{A\omega}+\rho_{B\omega} = A_0 + A_1^+\omega + A_2^+\omega^2 + \cdots\ ,
\end{equation}
where (up to second order)
%
\begin{eqnarray}
\label{eqa1}
A_0 & \equiv & \rho_{A,N_A}^{(0)}+\rho_{B,N_B}^{(0)}  \\
\label{eqa2}
A_1^+ & \equiv & \left|\phi_{A,N_A+1}^{(0)}\right|^2 - \left|\phi_{B,N_B}^{(0)}\right|^2 +\frac{\partial\left(\rho_{A,N_A}^{(\omega)} + \rho_{B,N_B}^{(\omega)}\right)}{\partial \omega}\Big|_{\omega=0^+}  \\
\label{eqa3}
A_2^+ &\equiv & 2\phi_{A,N_A+1}^{(0)}\frac{\partial\phi_{A,N_A+1}^{(\omega)}}{\partial \omega}\Big|_{\omega=0^+} - \\
\nonumber
&& - 2\phi_{B,N_B}^{(0)} \frac{\partial\phi_{B,N_B}^{(\omega)}}{\partial \omega}\Big|_{\omega=0^+} + \frac{\partial^2\left(\rho_{A,N_A}^{(\omega)}+ \rho_{B,N_B}^{(\omega)}\right)}{\partial \omega^2}\Big|_{\omega=0^+}\ ,
\end{eqnarray}
%
and we have assumed real single-particle orbitals.
The terms $A_1^+$, $A_2^+$ and higher include, through the
derivatives of the densities and orbitals, the 
linear, quadratic and higher response
of the embedded subsystems to a charge transfer $A\rightarrow B$.
These are thus very complex terms (see also subsection \ref{fukui_sec}).
Nevertheless, for the exact embedding we must have
$A_0=\rho^{KS}$, $A_i^+=0$ for $i\geq1$. 
These conditions state in fact the independence
of the total embedding density from the fractional occupation
of the subsystem densities for an exact embedding approach: 
removing a fraction of electrons from the $N_B$-th orbital of subsystem $B$ and
putting it on the $(N_A+1)$-th orbital of subsystem $A$ will
have no practical effects, since the variations of the density
induced by this ``charge transfer'' are compensated
order-by-order by equal and opposite relaxations of the
rest of the density induced by an appropriate embedding potential.

On the other hand, in the more general case of an
approximate embedding procedure the total 
embedding density will be no more independent from
the choice of $\omega$. 
In particular, it is well known that, using semilocal
kinetic energy approximations, non-negligible 
errors are found for systems with some charge-transfer character
\cite{gotz09,emb_ct}.
It is meaningful therefore to
consider the behavior as a function of $\omega$ of 
the error on the embedding density
\begin{eqnarray}
\nonumber
\Delta\rho_\omega^+ & = & \rho_{A\omega} + \rho_{B\omega} - \rho^{KS} \approx \\
\label{de}
& \approx & \Delta\rho_0  + A_1^+\omega + A_2^+\omega^2 + \cdots\ .
\end{eqnarray}
with $\Delta\rho_0=A_0-\rho^{KS}$.

\begin{figure}
\begin{center}
\includegraphics[width=\columnwidth]{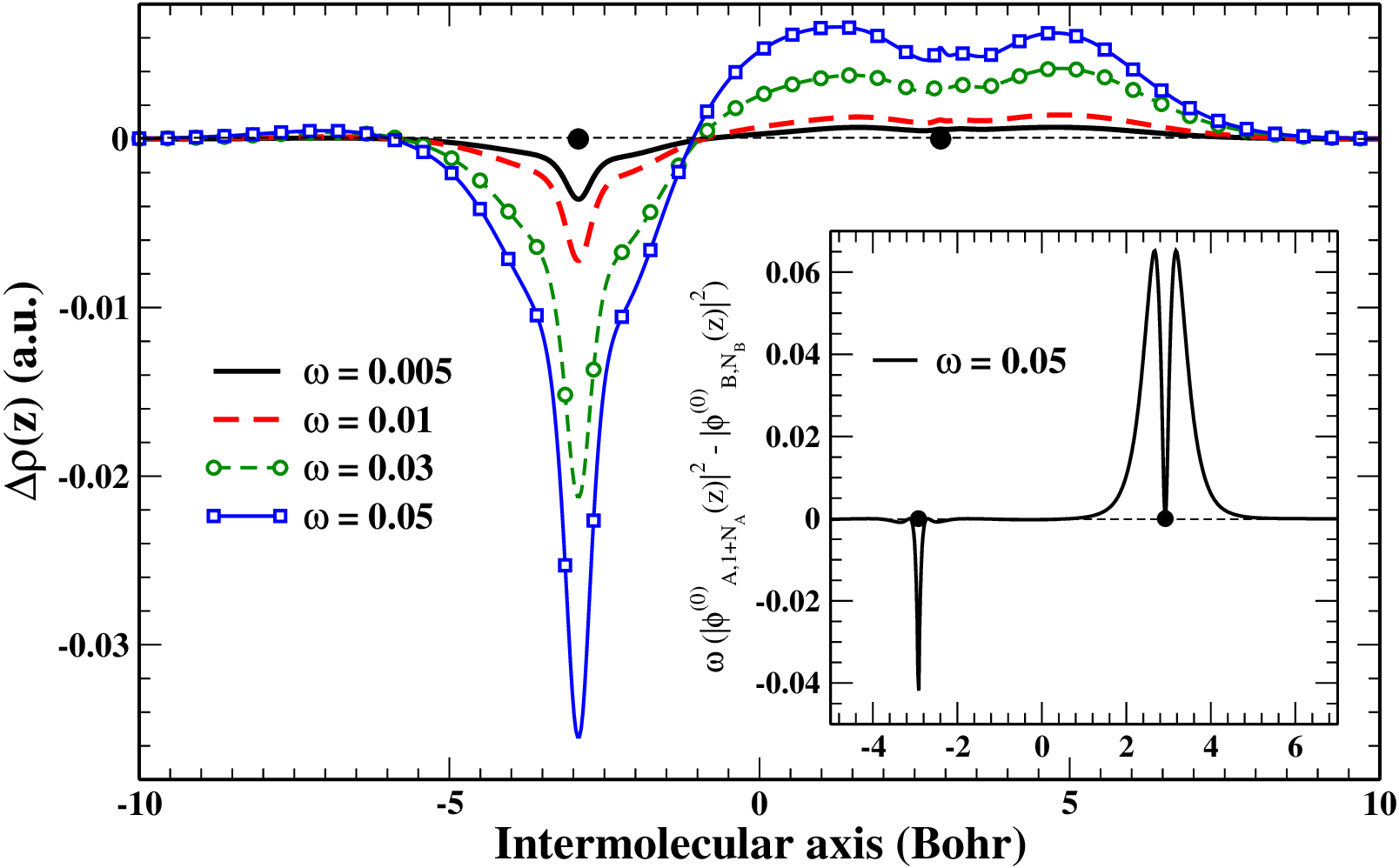}
\caption{Deviations of plane-averaged errors on embedding densities 
computed at different fractional particle numbers $\omega$ from the
reference one computed at $\omega=0$, for the Ne-Ne complex. 
The inset shows the frontier orbitals density difference
$|\phi_{A,N_A+1}^{(0)}|^2-|\phi_{B,N_B}^{(0)}|^2$.
The filled circles on the $x$-axis denote the atoms' positions}
\label{fig2}
\end{center}
\end{figure}
\begin{figure}
\begin{center}
\includegraphics[width=\columnwidth]{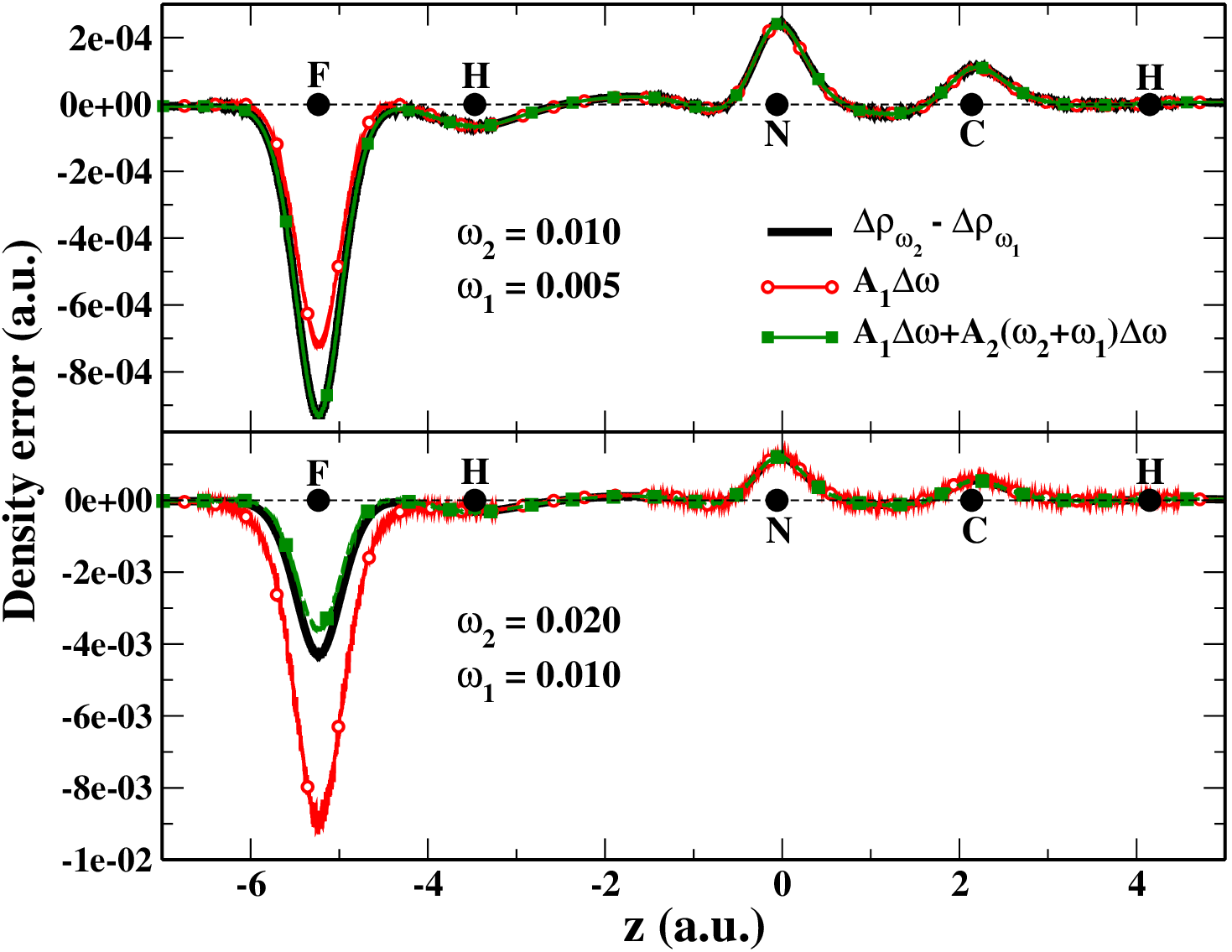}
\caption{\label{fig1} Differences between errors on embedding densities
computed using different values of the fractional occupation $\omega$
and comparison with the theoretical formula of Eq. (\ref{rrr}). 
The plot is on an axis parallel to the intermolecular axis of the HF-NCH
complex and displaced 0.5 a.u. from the atoms' plane.}
\end{center}
\end{figure}

Equation (\ref{de}) provides three important results.
First, the error on the embedding density can be expected to vary,
at each point of space, almost linearly near $\omega=0$
and roughly follow the spatial shape of 
$|\phi_{A,N_A+1}^{(0)}|^2-|\phi_{B,N_B}^{(0)}|^2$.
This fact is shown in Fig. \ref{fig2}, where we report 
$\Delta\rho_\omega^+$ for the Ne-Ne complex and compare it
to the shape of the frontier orbitals density difference
(in the inset).
The second result that can be deduced from Eq. (\ref{de}) is that
the difference between the errors on the embedding density 
computed at two different (small) values of the fractional charge
$\omega_1$ and $\omega_2$ is roughly proportional 
to $\Delta\omega=\omega_2-\omega_1$ (see Fig. \ref{fig1}). More precisely we have
\begin{equation}\label{rrr}
\Delta\rho_{\omega_2}^+ - \Delta\rho_{\omega_1}^+ = A_1^+\Delta\omega + A_2^+(\omega_2+\omega_1)\Delta\omega + \mathcal{O}\left((\omega_2+\omega_1)^2\right)\ .
\end{equation}
Finally, we see that
in order to minimize the error on the embedding density the
$\Delta\rho_0$ and $\omega A_1^+$ terms must cancel each other
(unless they are both zero, which however brings back to
the exact embedding case).
This situation is likely to occur if 
$\Delta \rho_0\approx\omega(|\phi_{A,N_A+1}^{(0)}|^2 - |\phi_{B,N_B}^{(0)}|^2)$
(the relaxation term is in general rather small).
Thus, the error will be minimized if the error
$\Delta\rho_0$ is dominated by
a (small) charge-transfer of $\omega$ electrons from the subsystem A
to the subsystem B.
This suggests that the error on the embedding density
is minimized by non-zero values of $\omega$ for charge-transfer
complexes, whereas it is minimum close to $\omega=0$ for other
types of non-covalent complexes (e.g. hydrogen bond, dipole-dipole,
or dispersion complexes). 

\begin{figure}
\begin{center}
\includegraphics[width=\columnwidth]{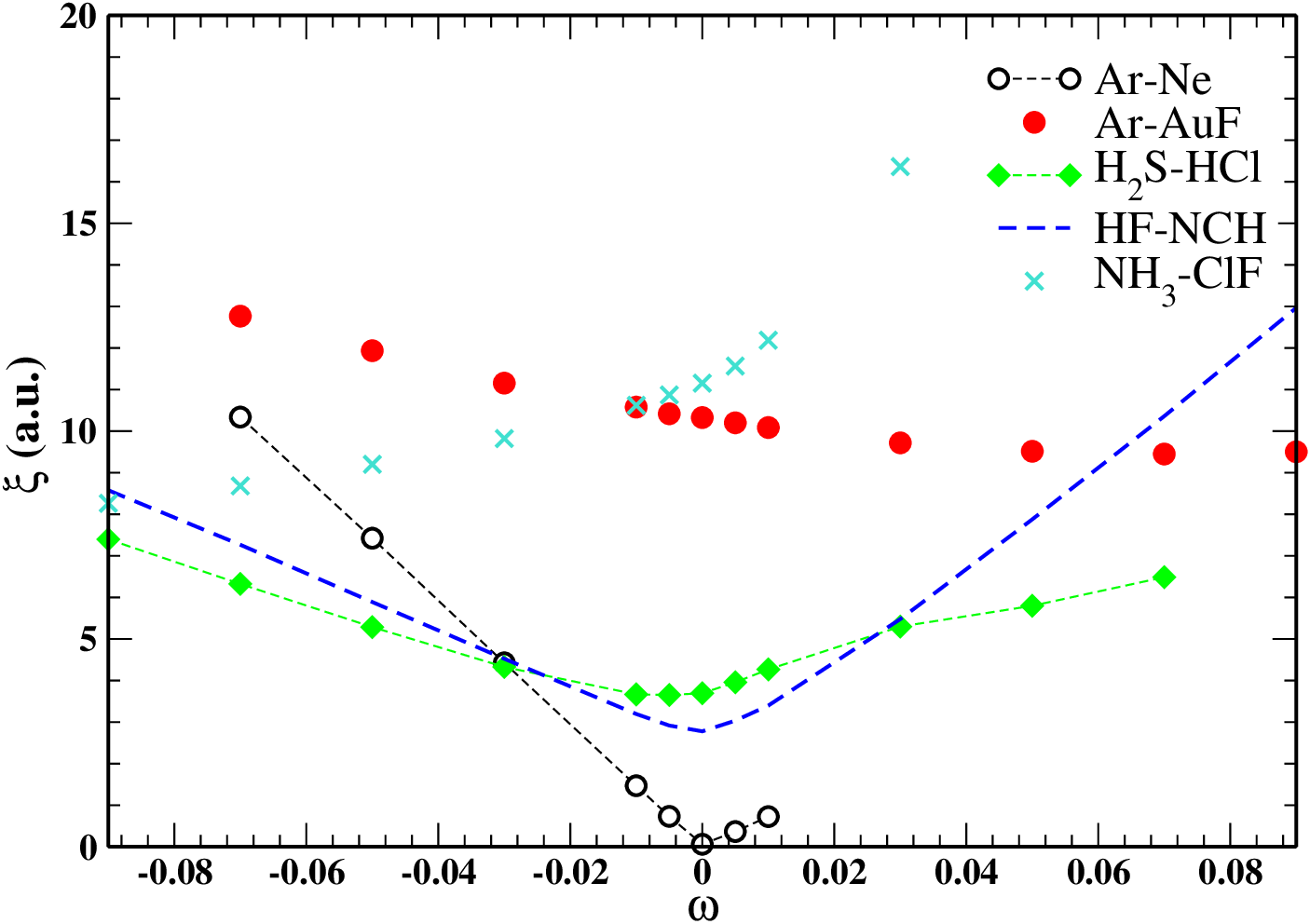}
\caption{Embedding density errors for all the 
systems analyzed in this work using different values of the fractional particle 
number $\omega$.}
\label{fig3}
\end{center}
\end{figure}

This result finds support in
Fig. \ref{fig3} where we report the integrated error on
embedding density
\begin{equation}
\xi_\omega = \frac{1000}{N_A+N_B}\int \left|\Delta\rho_\omega^\pm(\R)\right|d\R
\end{equation}
for various non-covalent complexes. Inspection of the plots
shows in fact that, as expected, the error is minimized
at $\omega=0$ for the Ar-Ne, H$_2$S-HCl, and HF-NCH complexes, whereas 
for the charge-transfer complexes Ar-AuF and NH$_3$-ClF the minimum
of $\xi_\omega$ is found for $\omega\neq0$. 
Note also the quasi linear behavior of the error for different
values of $\omega$, in agreement with the previous discussion. 

To conclude this section we add that Eq. (\ref{de}) can be
finally used to obtain some bounds on the absolute error
on embedding density, so that 
\begin{equation}
\left||\Delta\rho_0|\mp\omega|A_1^\pm|\right| \lessapprox \left|\Delta\rho_\omega^\pm\right| \lessapprox \left|\Delta\rho_0\right| \pm \omega |A_1^\pm|\ .
\end{equation}

\subsection{Comments on the embedded Fukui function}
\label{fukui_sec}
The Fukui function \cite{geerlings03,fukui_chap} 
is an important chemical descriptor
defined as the differential change in electron density due to an
infinitesimal variation in the particle number. 
Its meaning is to identify those regions
of space where a system is more susceptible of
nucleophilic or electrophilic attack, i.e. where it
can better accommodate a positive or negative change in the
electron number.
According to these definitions, in the present case we can 
define four distinct embedded Fukui functions
\begin{eqnarray}
\label{fuk1}
f^+_A(\R) \equiv \frac{\partial\rho_{A\omega}(\R)}{\partial \omega}\Big|_{\omega=0^+} & \ ; \ & f^-_B(\R) \equiv -\frac{\partial\rho_{B\omega}(\R)}{\partial \omega}\Big|_{\omega=0^+} \\
\label{fuk2}
f^-_A(\R) \equiv \frac{\partial\rho_{A\omega}(\R)}{\partial \omega}\Big|_{\omega=0^-} & \ ; \ & f^+_B(\R) \equiv -\frac{\partial\rho_{B\omega}(\R)}{\partial \omega}\Big|_{\omega=0^-}
\end{eqnarray}
In fact, Eqs. (\ref{fuk1}) and (\ref{fuk2}) imply that
large values of $f_A^+$ in one region of space
provide an indication that the embedded subsystem 
A is able to easily accommodate excess charge
in that region, whereas large values of $f_A^-$ 
indicate that some charge can be easily removed
from a certain region of the embedded subsystem A. 
Similar considerations hold for subsystem B.
We note that the Fukui functions is Eqs. 
(\ref{fuk1}) and (\ref{fuk2}) are defined for $\omega=0^+/0^-$.
This is the most natural definition, because  
i) in approximated embedding calculations $\omega$=0 
gives very often the best results in terms of energy 
and density (note that the current kinetic-energy 
functionals fail to describe long-range charge-transfer 
systems), and ii) it can be directly related to the the 
case of isolated subsystems.

Moreover, we note that, whenever the subsystem densities are obtained with
a full self-consistent procedure, the definitions
(\ref{fuk1}) and (\ref{fuk2}) present no ambiguity,
as the subsystem densities are uniquely defined
(using a freeze-and-thaw scheme 
no arbitrary choice is required for the frozen densities) at any fixed $\omega$.
Moreover, because they are obtained within an
embedding approach, the resulting Fukui
functions fully include all the
environmental effects.

The Fukui functions defined in Eqs. (\ref{fuk1}) and (\ref{fuk2})
can be further manipulated to obtain important expressions.
In order to do this we consider in the following
$f_A^+$; similar derivations can be carried out for
all other cases.
Using Eq. (\ref{e27}), we can write
\begin{eqnarray}
\nonumber
\frac{\partial\rho_{A\omega}}{\partial \omega} & = & \rho_{A,N_A+1}^{(\omega)} + \omega\frac{\partial \rho_{A,N_A+1}^{(\omega)}}{\partial \omega} - \\
\nonumber
&& - \rho_{A,N_A}^{(\omega)} + (1-\omega)\frac{\partial \rho_{A,N_A}^{(\omega)}}{\partial \omega} = \\
\nonumber
& = &\left|\phi_{A,N_A+1}^{(\omega)}\right|^2 + \frac{\partial \rho_{A,N_A}^{(\omega)}}{\partial \omega} + \\
\label{jtu}
&& + \omega\left[\frac{\partial\rho_{A,N_A+1}^{(\omega)}}{\partial \omega} - \frac{\partial\rho_{A,N_A}^{(\omega)}}{\partial \omega}\right]\ .
\end{eqnarray}
Hence, in the limit $\omega\rightarrow 0^+$ we have that the Fukui
  function is given by the frontier orbital density plus a relaxation term, in
  agreement with the analogous result obtained via the partition
  density-functional theory \cite{cohen07}.

To gain more inside into Eq. (\ref{jtu}) we consider however, that
the partial derivatives are computed, using the chain rule, as \cite{hellgren12_2}
\begin{eqnarray}
\nonumber
\frac{\partial\rho_{A,M}^{(\omega)}(\R)}{\partial \omega} & = & \int \frac{\delta\rho_{A,M}^{(\omega)}(\R)}{\delta v^A_{s+e}(\R_1)}\frac{\delta v^A_{s+e}(\R_1)}{\delta\rho_{A\omega}(\R_2)}\frac{\partial\rho_{A\omega}(\R_2)}{\partial\omega} d\R_1d\R_2= \\
\nonumber
& = & \int \chi_{A,M}(\R,\R_1)K_{Hxce}^{A+}(\R_1,\R_2)\frac{\partial\rho_{A\omega}(\R_2)}{\partial\omega}d\R_1d\R_2 \ ,\\
\label{cbvf}
\end{eqnarray}
where $v^A_{s+e}=v_s[\rho_{A\omega}]+v_{emb}^A$, $\chi_{A,M}$ is the linear 
response function of the $M$-particle subsystem A 
(see Ref. \onlinecite{pavanello13} for details on $\chi_{A,M}$), 
and $K_{Hxce}^{A+}$ is the Hartree-XC-embedding kernel 
relative to subsystem A in the case $\omega\geq 0$ 
(we remark that the XC kernel is discontinuous 
at integer particle numbers \cite{hellgren12,hellgren12_2}).
We note that in Eq. (\ref{cbvf}) all the derivatives 
are well defined: in fact, $\rho_{A,M}^{(\omega)}$
is built from $M$ KS orbitals that are all 
determined by the potential  $v^A_{s+e}=v_s[\rho_{A\omega}]+v_{emb}^A$
(see Eq. (\ref{e21})); on the other hand, $v^A_{s+e}$ 
is clearly a functional of $\rho_{A\omega}$.
Then, using Eq. (\ref{cbvf}) into Eq. (\ref{jtu}) and taking the limit
$\omega\rightarrow0^+$ we finally obtain
\begin{eqnarray}
\nonumber
&&f_A^+(\R) = \left|\phi_{A,N_A+1}^{(0)}(\R)\right|^2 +\\
&&\quad + \int \chi_{A,N_A}(\R,\R_1)K_{Hxce}^{A+}(\R_1,\R_2)f_A^+(\R_2) d\R_1d\R_2\ .
\end{eqnarray}
This expression shows that, in first approximation, 
the effect of the environment on the Fukui function
is simply accounted for by a perturbation induced by the 
embedding potential on the frontier molecular orbital of the isolated subsystem A.
However, the second term includes more subtle effects,
since it incorporates not only the response of the
subsystem A to the perturbation (the change in the
particle number), but also the response of the whole
environment to it. In fact, the $K_{Hxce}^{A+}$ kernel
contains the functional derivative with respect to $\rho_{A\omega}$ of
the embedding potential, which is a bifunctional of both $\rho_{A\omega}$
and $\rho_{B\omega}$ (see Eq. (\ref{e22})). Thus, it contains also
contributions proportional to $\delta\rho_{B\omega}/\delta\rho_{A\omega}$, which
describe the relaxation of the environment to the
nucleophilic attack on the subsystem A.

The fact that the embedded Fukui functions include
the effects of the whole system has important consequences.
The most important is that (at least for exact embedding)
we have an equilibration principle for the embedded Fukui 
function. In fact, it is easy to show that 
because $\rho_{A\omega}+\rho_{B\omega}=\rho$ and $\rho$ is independent
on $\omega$, we must have at any point in space 
$f_A^+=f_B^-$ (and $f_A^-=f_B^+$).
The rationale beyond this finding is that the embedding
is describing a ground state solution
for the whole system. Thus, the propensity to acquire and
donate a certain amount of charge must be equal in the whole space
to guarantee the stability.
This result nicely agrees with that of Eq. (4.28) of Ref. \cite{cohen07}.

\section{Effect of the non-integer particle numbers on the embedding energy}
According to Eq. (\ref{eq:totE}) the total energy 
resulting from the embedding procedure is
\begin{equation}\label{aaa1}
E^{emb}[\rho_{A\omega},\rho_{B\omega}]= E[\rho_{A\omega}]+ E[\rho_{B\omega}]+\mathcal{E}^{nadd}[\rho_{A\omega};\rho_{B\omega}]
\end{equation}
with
\begin{eqnarray}
\nonumber
&&E[\rho_{j}]  =  T_s[\rho_{j}]+ \int\rho_{j}(\R) v^j_{ext}(\R)\, d\R +J[\rho_{j}]+E_{xc}[\rho_{j}] \\
\label{aaa2}
&& \\
\nonumber
&&\mathcal{E}^{nadd}[\rho_{A\omega};\rho_{B\omega}]  =  \int\frac{\rho_{A\omega}(\R_1)\rho_{B\omega}(\R_2)}{\left|\R_1-\R_2\right|}d\R_1d\R_2 + \\
\nonumber
&&\ \ \ \ +  \int\rho_{B\omega}(\R) v^A_{ext}(\R)d\R + \int\rho_{A\omega}(\R) v^B_{ext}(\R)d\R +\\
\label{enadd}
&& \ \ \ \ +  T_s^{nadd}[\rho_{A\omega};\rho_{B\omega}]+ E_{xc}^{nadd}[\rho_{A\omega};\rho_{B\omega}]\ ,
\end{eqnarray}
where $j=A,B$.
For convenience we will call these the subsystems' and the
non-additive contributions to the total energy, respectively.
The variation of this energy with respect to $\omega$ can be computed 
by means of the chain rule 
$\partial/\partial\omega = \int(\delta/\delta\rho_j(\R_1))(\partial\rho_j(\R_1)/\partial\omega)d\R_1$.
Thus, after some algebra (see Appendix \ref{appa})
we obtain
\begin{equation}\label{dedw}
\left(\frac{\partial E^{emb}}{\partial \omega}\right)^\pm = \mu_A^\pm-\mu_B^\pm\ ,
\end{equation}
where the $+$ and $-$ superscripts indicate the $\omega>0$ 
and $\omega<0$ cases, respectively, and
$\mu_A$ and $\mu_B$ are the chemical potentials of the 
embedded subsystems $A$ and $B$, respectively.

Equation (\ref{dedw}) states the simple physical fact that
the total embedding energy
change upon a transfer of charge $\delta\omega$ from one
subsystem to the other is proportional to
the difference of the chemical potentials of
the two subsystems. This result is analogous to
the well known expression for the formation energy
in the HSAB theory \cite{chermette99}.

Of course, for an exact embedding calculation
$\mu_A^\pm=\mu_B^\pm$ by definition. Thus, the
total embedding energy is independent from
the fractional charge $\omega$, as it must be.
However, for an approximate embedding calculation
we can expect the equilibration of the chemical
potentials to be incomplete \cite{gritsenko_chap}. Then, we can 
investigate the behavior of the embedding energy error
$\Delta E^{emb}_\omega=E^{emb}-E_0$, where $E_0$ is the
exact KS total energy of the total system, which is independent on $\omega$.
Hence, we can consider
\begin{equation}
\left(\frac{\partial \Delta E^{emb}_\omega}{\partial \omega}\right)^\pm = \mu_A^\pm-\mu_B^\pm\ .
\end{equation}

To evaluate the chemical potentials we 
cannot make direct use of the Janak's theorem
\cite{perdew82,janak78}, because the
energies of the individual interacting subsystems
are not well defined in the FDE formalism.
Nevertheless, the chemical potentials can be
evaluated considering that at each
value of $\omega$, despite the energies of the
interacting and non-interacting Kohn-Sham systems
are different, their chemical potentials are
equal by definition. Therefore, focusing, for
the moment, on $\mu_A^+$, we have
\begin{equation}\label{oiu}
\mu_A^+ = \left(\frac{\partial E_{s,0}^{A+}}{\partial\omega}\right)_{\rho_{A\omega}}\ ,
\end{equation}
where 
\begin{equation}\label{uyt}
E_{s,0}^{A+} = \Tr{\hat{H}\hat{\Gamma}_A^{KS}} = \omega E_{A,N_A+1}^{KS} + (1-\omega)E_{A,N_A}^{KS}\ ,
\end{equation}
is the energy of the non-interacting embedded 
Kohn-Sham system (for $\omega>0$), with
$\hat{H}$ being the KS Hamiltonian,
\begin{equation}\label{rew}
E_{A,M}^{KS} = \sum_i^M\epsilon_{A,i}^{(\omega)}\ ,
\end{equation}
and $\epsilon_{A,i}^{(\omega)}$ being the eigenvalues 
of the KSCED for subsystem A.
The subscript $\rho_{A\omega}$ recalls that the derivative must be evaluated 
keeping $\rho_{A\omega}$ fixed, because the equivalence of the Kohn-Sham and the
interacting systems only holds at the ground state density (note 
that this requirement is equivalent to the constraint
of fixing the external nuclear potential used in conventional calculations
of the chemical potential). However, the embedding potential,
depending also on $\rho_{B\omega}$, is allowed to vary.

From Eqs. (\ref{oiu}), (\ref{uyt}), and (\ref{rew}) we readily find
\begin{equation}\label{cvf}
\mu_A^+ = \epsilon_{A,N_A+1}^{(\omega)} + \sum_i^{N_A+1}f_i\left(\frac{\partial\epsilon_{A,i}^{(\omega)}}{\partial\omega}\right)_{\rho_{A\omega}} \approx \epsilon_{A,N_A+1}^{(\omega)}\ ,
\end{equation}
where $f_i$ are the occupation numbers.
The term depending on the derivatives of the orbital energies
is a relaxation term, expressing the response of the system
to changes in the embedding potential due to $\omega$-induced variations
of $\rho_{B\omega}$. In fact, we can write
\begin{equation}
\left(\frac{\partial\epsilon_{A,i}^{(\omega)}}{\partial\omega}\right)_{\rho_{A\omega}} = \int \frac{\delta\epsilon_{A,i}^{(\omega)}}{\delta v_{emb}^A(\R_1)}\frac{\delta  v_{emb}^A(\R_1)}{\delta\rho_{B\omega}(\R_2)}\frac{\partial \rho_{B\omega}(\R_2)}{\partial\omega}d\R_1d\R_2\ .
\end{equation}
This term is generally small with respect to the frontier
orbital energy and was neglected in the last equality of Eq. (\ref{cvf}).

Then, summarizing we find
\begin{eqnarray}
\label{dembdw1}
\left(\frac{\partial\Delta E^{emb}_\omega }{\partial \omega}\right)^+ & \approx & \epsilon_{A,N_A+1}^{(\omega)} - \epsilon_{B,N_B}^{(\omega)} \\ 
\label{dembdw2}
\left(\frac{\partial \Delta E^{emb}_\omega}{\partial \omega}\right)^- & \approx & \epsilon_{A,N_A}^{(\omega)} - \epsilon_{B,N_B+1}^{(\omega)} \ .
\end{eqnarray}
These equations formalize the intuitive idea
that, apart for relaxation contributions,
the energy change due to the transfer of
a fraction of electron from one subsystem to the other 
is proportional to the difference
of the frontier 
(i.e. with fractional occupation) orbitals eigenvalues.

\begin{figure}
\begin{center}
\includegraphics[width=\columnwidth]{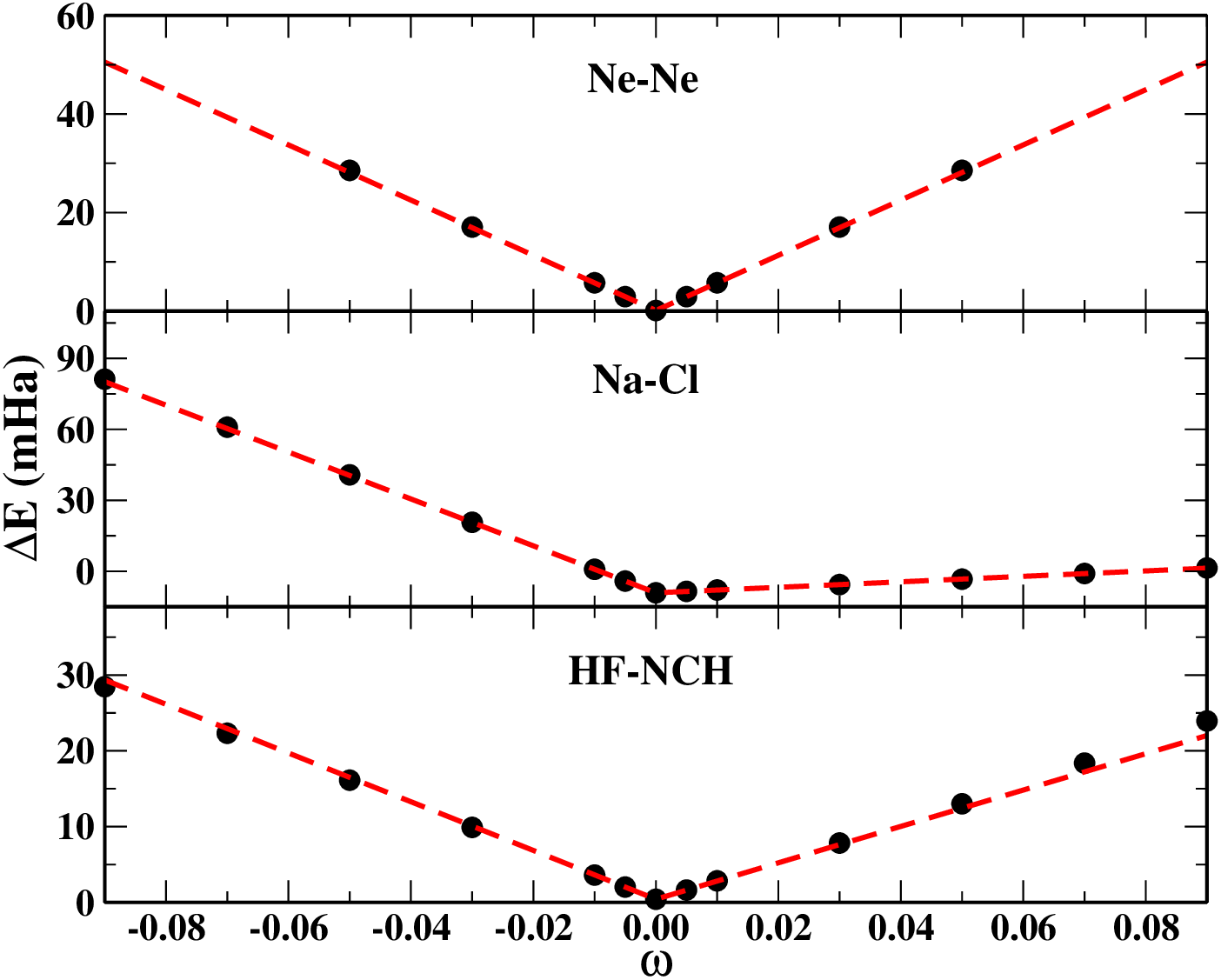}
\caption{Embedding energy errors $\Delta E_\omega^{emb}$ 
(filled circles) at different values of the fractional 
occupation $\omega$. Dashed lines indicate theoretical fits 
according to Eqs. (\ref{dembdw1}) and (\ref{dembdw2}).
In all calculations we used revAPBEk functional for the non-additive
kinetic energy and a supermolecular def2-TZVPPD basis set. All values reported
in these plots are in mHa.}
\label{fig4}
\end{center}
\end{figure}

Equations (\ref{dembdw1}) and (\ref{dembdw2}) are used in Fig. \ref{fig4}
to plot the theoretical behavior (red dashed-lines) of $\partial\Delta E^{emb}_\omega/\partial \omega$
for several non-covalent complexes 
and compare it to numerical results (black dots) from 
embedding calculations at different $\omega$. 
Fig. \ref{fig4} shows that very large errors (10-50 mHa) are present in the energy already for very small $\omega$
values.
(at $\omega=0$ the errors are 0.14, -9.07 and 0.43 mHa, for Ne-Ne, NaCl and HF-NCH, respectively).
This behavior is related to the incorrect kinetic energy potential which largely fails to reproduce the
total density (see e.g. Fig. \ref{fig1} for the Ne dimer) and the chemical potentials of embedded subsystems.
The observed strong asymmetry for positive and negative $\omega$ values in the plot of Na-Cl 
(partitioned as Na$^+$ and Cl$^-$) is related to the asymmetric energy levels of the embedded subsystems. 
In particular, when $\omega>0$ we add electrons to the  LUMO of Na$^+$ (at about -2 eV) and we 
remove electrons from the HOMO of Cl$^-$ (at about -5 eV); on the other 
hand, when $\omega<0$ we remove electrons from the HOMO of Na$^+$ (at about -29 eV) and add electrons to the 
LUMO of Cl$^-$ (at about -2 eV). Thus, only for $\omega>0$ the chemical potentials (i.e. the fractional 
occupied orbitals) are almost aligned, which corresponds to lower slopes through Eqs. (60) and (61).

We acknowledge that similar plots for the behavior of the total energy as a function of the
subsystems’ occupations were also reported within PDFT \cite{tang12}. In both cases the cusp 
originates from the inability of the embedding/partition potential to correctly align the subsystems' 
chemical potentials (i.e. the orbitals with fractional occupation). However, the reasons for this 
drawback are different in the two cases because of the basic theoretical differences between the 
two methods. In fact, in our FDE treatment the embedding potential is incorrect (for every occupation number) 
due to the approximations in the kinetic energy functional, while in Ref. \onlinecite{tang12} 
the embedding potential is 
exact (by construction) only for the “true” occupation numbers.

\subsection{Comments on the derivative discontinuity}
Equations (\ref{dembdw1}) and (\ref{dembdw2}) establish
a relation between the variation of the embedding total energy
(or equivalently of the error on the embedding energy) with
the fractional charge and the difference in the 
frontier orbital energies of the two subsystems.
However, for an exact embedding calculation the
embedding energy must be independent from the fractional
charge. Thus, we must have
\begin{eqnarray}
\epsilon_{A,N_A+1}^{(\omega)}  \approx  \epsilon_{B,N_B}^{(\omega)} & \quad & \mathrm{for}\ \ \omega>0 \\
\epsilon_{A,N_A}^{(\omega)} \approx \epsilon_{B,N_B+1}^{(\omega)} & \quad & \mathrm{for}\ \ \omega<0\ .
\end{eqnarray}
These equations are rationalized in terms
of the need for the two subsystems to align their
frontier orbital levels in order to achieve
electrochemical equilibrium. This behavior is independent
on the amount of charge transferred, provided that
$\omega\neq0$.
On the other hand, no such need arises when 
$\omega=0$. In fact, in this case we
rather would have 
\begin{equation}
\frac{\epsilon_{A,N_A+1}^{(0)}+\epsilon_{A,N_A}^{(0)}}{2}\approx \frac{\epsilon_{B,N_B+1}^{(0)}+\epsilon_{B,N_B}^{(0)}}{2}\ .
\end{equation}
Thus, a clear discontinuity can be observed in the
system when the fractional charge is turned on.

This discontinuous behavior is triggered by a discontinuity 
in the embedding potential at integer subsystems' occupations.
The effect of a discontinuity in the potential is in fact 
to promote a sudden shift in the orbital energies.
On the other hand, the effect cannot be ascribed to
the discontinuity in the Kohn-Sham potential and
related kinetic energy, because these already act
on the isolated subsystems. Thus, the whole
shift required for equilibration must be 
related to a derivative discontinuity of the
non-additive embedding contributions.

Using Eq. (\ref{enadd}) indeed we find
\begin{eqnarray}
\nonumber
v_{emb}^{A\pm} & = & \left(\frac{\delta\mathcal{E}^{nadd}}{\delta\rho_{A\omega}}\right)^\pm = \frac{\delta T_s[\rho_{A\omega}+\rho_{B\omega}]}{\delta (\rho_{A\omega}+\rho_{B\omega})} - \\
\nonumber
&& - \left(\frac{\delta T_s[\rho_{A\omega}]}{\delta \rho_{A\omega}}\right)^\pm + \left(\frac{\delta E_{xc}[\rho_{A\omega}+\rho_{B\omega}]}{\delta (\rho_{A\omega}+\rho_{B\omega})}\right)^\pm -\\
&& - \left(\frac{\delta E_{xc}[\rho_{A\omega}]}{\delta \rho_{A\omega}}\right)^\pm + v_J^B + v_{ext}^B \\
\nonumber
v_{emb}^{B\pm} & = & \left(\frac{\delta\mathcal{E}^{nadd}}{\delta\rho_{B\omega}}\right)^\pm = \frac{\delta T_s[\rho_{A\omega}+\rho_{B\omega}]}{\delta (\rho_{A\omega}+\rho_{B\omega})} - \\
\nonumber
&& - \left(\frac{\delta T_s[\rho_{B\omega}]}{\delta \rho_{B\omega}}\right)^\pm + \left(\frac{\delta E_{xc}[\rho_{A\omega}+\rho_{B\omega}]}{\delta (\rho_{A\omega}+\rho_{B\omega})}\right)^\pm -\\
&& - \left(\frac{\delta E_{xc}[\rho_{B\omega}]}{\delta \rho_{B\omega}}\right)^\pm + v_J^A + v_{ext}^A\ ,
\end{eqnarray}
where we used the fact that 
$\delta (\rho_{A\omega}(\R_1)+\rho_{B\omega}(\R_1))/\delta \rho_i(\R_2)=\delta(\R_1-\R_2)$.
Note that the $\pm$ superscript was used neither for
$\delta T_s[\rho_{A\omega}+\rho_{B\omega}]/\delta(\rho_{A\omega}+\rho_{B\omega})$, 
since this derivative is 
computed for the integer density $\rho_{A\omega}+\rho_{B\omega}$, nor for
$v_J$ and $v_{ext}$ that are continuous functionals of the density.
As a consequence, the discontinuity of the
embedding potential is found to be
\begin{eqnarray}
\nonumber
\Delta_{emb}^A & \equiv & v_{emb}^{A+}-v_{emb}^{A-} = \left(\frac{\delta T_s[\rho_{A\omega}]}{\delta \rho_{A\omega}}\right)^- - \left(\frac{\delta T_s[\rho_{A\omega}]}{\delta \rho_{A\omega}}\right)^+ + \\
&&+ \left(\frac{\delta E_{xc}[\rho_{A\omega}]}{\delta \rho_{A\omega}}\right)^- - \left(\frac{\delta E_{xc}[\rho_{A\omega}]}{\delta \rho_{A\omega}}\right)^+ \\
\nonumber
\Delta_{emb}^B & \equiv & v_{emb}^{B+}-v_{emb}^{B-} = \left(\frac{\delta T_s[\rho_{B\omega}]}{\delta \rho_{B\omega}}\right)^- - \left(\frac{\delta T_s[\rho_{B\omega}]}{\delta \rho_{B\omega}}\right)^+ +  \\
&& + \left(\frac{\delta E_{xc}[\rho_{B\omega}]}{\delta \rho_{B\omega}}\right)^- - \left(\frac{\delta E_{xc}[\rho_{B\omega}]}{\delta \rho_{B\omega}}\right)^+\ .
\end{eqnarray}
These discontinuities are exactly opposite
to those due to the KS potential and the 
non-interacting kinetic energy in each
(isolated) subsystem
\begin{eqnarray}
\nonumber
\Delta_s^{A} & = & \left(\frac{\delta T_s[\rho_{A\omega}]}{\delta \rho_{A\omega}}\right)^+ - \left(\frac{\delta T_s[\rho_{A\omega}]}{\delta \rho_{A\omega}}\right)^- + \\
&&+ \left(\frac{\delta E_{xc}[\rho_{A\omega}]}{\delta \rho_{A\omega}}\right)^+ - \left(\frac{\delta E_{xc}[\rho_{A\omega}]}{\delta \rho_{A\omega}}\right)^- \\
\nonumber
\Delta_s^{B} & = & \left(\frac{\delta T_s[\rho_{B\omega}]}{\delta \rho_{B\omega}}\right)^+ - \left(\frac{\delta T_s[\rho_{B\omega}]}{\delta \rho_{B\omega}}\right)^- +  \\
&& + \left(\frac{\delta E_{xc}[\rho_{B\omega}]}{\delta \rho_{B\omega}}\right)^+ - \left(\frac{\delta E_{xc}[\rho_{B\omega}]}{\delta \rho_{B\omega}}\right)^-\ .
\end{eqnarray}

Thus, two main cases can be considered.
(1) Both the exchange-correlation and the
kinetic energy are treated in the same way
(eventually exactly)
in the subsystems' and the non-additive contributions
to the total energy. This corresponds
to an exact embedding treatment.
In this case the discontinuities cancel each
other and the derivative of the 
total embedding energy is continuous
at integer subsystems' particle numbers. In fact,
the embedding energy is even constant, because the 
discontinuity in the embedding potential is such as to 
force the equilibration of the subsystems' chemical
potentials at any $\omega$.
(2) The exchange-correlation energy is treated
in the same way both in the subsystems' and the 
non-additive contributions
to the total energy but the kinetic energy is not.
Typically, in a KS framework we would have
that the non-interacting kinetic energy is exact
in the subsystems' contributions to the total energy
but approximate by a semilocal functional
in the non-additive one.
This is the usual case in practical embedding
calculations. In this situation the total
exchange-correlation discontinuity in $\Delta_{emb}^i+\Delta_s^i$ 
cancels out, however there is no discontinuity
in the embedding kinetic term of $\Delta_{emb}^i$
because a semilocal kinetic approximation is used.
Thus, the KSCED equations include
a spurious kinetic derivative discontinuity.
As a result the derivative of the
total embedding energy displays
a cusp at integer subsystems' particle numbers.
This cusp is determined by the kinetic
derivative discontinuity and therefore is 
equal to the frontier orbitals
eigenvalues difference \cite{yang12,sagvolden08}
, in agreement with Eqs. 
(\ref{dembdw1}) and (\ref{dembdw2}).

\subsection{Comments on the embedded chemical hardness}
The global hardness is defined as the derivative of the 
chemical potential with respect to the particles number
at fixed nuclear environment \cite{parr_book,crt_book,chermette99,geerlings03}.
Because the nuclear potential is piecewise constant,
at zero temperature this definition is not well posed,
since the forward and backward derivatives are always zero
while the central derivative diverges.
Thus, in practice, at zero temperature, the original
definition of the hardness is replaced
by the difference of the chemical potentials computed 
at $N+\delta$ and $N-\delta$, respectively, with $\delta\rightarrow0^+$.

Thus, we can define the embedded subsystems' hardnesses as
\begin{equation}\label{hard}
\eta_A \equiv \mu_A^+ - \mu_A^- \quad ; \quad \eta_B \equiv - (\mu_B^+ - \mu_B^-)\ .
\end{equation}
These definitions respect the general features 
associated to the concept of chemical hardness \cite{parr_book}.
In particular, the fact that a small hardness corresponds to
the possibility to easy add or remove charge 
from the (embedded) system,
whereas a high hardness implies the opposite.
Equations (\ref{hard}) include into the hardness of
each subsystem the environmental effects (due to
the other subsystem), thus they provide
an indicator for the possibility that a component
of a complex may interact with a third system
(e.g., we can obtain the hardness
of an acidic complex when solvated and compare
it to that of a basic complex immersed in the
same solution).

Using Eq. (\ref{cvf}), the embedded hardness can be
easily calculated to be
\begin{eqnarray}
\eta_A & = & \epsilon_{A,N_A+1}^{(0)} - \epsilon_{A,N_A}^{(0)} + \\
\nonumber
&& + \sum_{i}^{N_A+1}\left(\frac{\partial\epsilon_{A,i}^{(0)}}{\partial\omega}\right)_{\rho_{A\omega}}^+ - \sum_{i}^{N_A}\left(\frac{\partial\epsilon_{A,i}^{(0)}}{\partial\omega}\right)_{\rho_{A\omega}}^- \\
\eta_B & = & \epsilon_{B,N_B+1}^{(0)} - \epsilon_{B,N_B}^{(0)} +  \\
\nonumber
&& + \sum_{i}^{N_B+1}\left(\frac{\partial\epsilon_{B,i}^{(0)}}{\partial\omega}\right)_{\rho_{B\omega}}^+ - \sum_{i}^{N_B}\left(\frac{\partial\epsilon_{B,i}^{(0)}}{\partial\omega}\right)_{\rho_{B\omega}}^-\ ,
\end{eqnarray}
where we took the limits $\omega\rightarrow0^+$ and
$\omega\rightarrow 0^-$ as appropriate.
As we observed for the chemical potential the relaxation
terms are usually small. Moreover, the two relaxation contributions
tend to cancel each other, apart for differences originating in the
discontinuities of the embedding potential derivative.
Thus, it may be appropriate in most situations to 
consider the approximate expressions
\begin{equation}\label{hh}
\eta_A \approx \epsilon_{A,N_A+1}^{(0)} - \epsilon_{A,N_A}^{(0)}\quad ; \quad \eta_B \approx \epsilon_{B,N_B+1}^{(0)} - \epsilon_{B,N_B}^{(0)}\ .
\end{equation}

\section{Computational details}
To test some of the results of theoretical analysis of the previous sections
FDE calculations were performed on a set of representative 
non-covalent complexes (Ne-Ne, Ne-Ar, Ar-AuF, H$_2$S-HCl, HF-NCH, (NH$_3$)$_2$, NH$_3$-ClF and NaCl).
Note that NaCl is partitioned as Na$^+$/Cl$^-$.
The structures of the complexes were taken from
Refs. \onlinecite{zhao05,zhao05_2,beyhan10,wesolowski96_2}.

The test set was restricted to contain only non-covalently bound
systems because these are the only ones that can be properly
described in FDE calculations at the current state of the art.
On the other hand, strongly interacting complexes, despite being 
potentially interesting systems to be studied with FDE calculations using
fractional subsystems' occupations, were not considered in the present
work to keep the analysis and the discussion more clear.

All calculations were performed with a development version of the
\texttt{FDE} script \cite{hyb_emb}, interfaced with the
\texttt{TURBOMOLE} program package \cite{TURBOMOLE}, version 6.4.
In all cases a freeze-and-thaw procedure \cite{wesolowski96} was used
to guarantee the full relaxation of the embedded ground-state electron density 
of both subsystems, until dipole moments of the embedded subsystems 
converged to $10^{-3}$ au. A supermolecular def2-TZVPPD \cite{weigend03,weigend05} 
basis set was employed in all calculations to expand the subsystem 
electron densities. 
The def2-TZVPPD basis set adds diffuse basis functions to 
the def2-TZVPP \cite{weigend05} basis set, thus granting an accurate 
representation for the electron densities even at the relatively 
large bonding distances characteristic of the systems under consideration.
Very accurate integration grids were employed to minimize numerical errors.
Additional details about our implementation and computational procedure 
are reported in Refs. \onlinecite{hyb_emb,emb_energy}.
The calculations were performed using the semilocal PBE XC functional
\cite{pbe}, while the revAPBEk \cite{apbek,apbe,apbekroutine} 
functional was used to 
approximate the non-additive non-interacting kinetic energy term.

\section{Conclusions}
In this paper we extended the FDE theory to the treatment
of non-integer subsystems' particle numbers and discussed the
relevance of this extension for the outcome of 
practical frozen-density simulations. 
In particular, we were able to show
that, in approximate 
FDE
applications,
the resulting densities and total energies have a well defined
behavior as functions of the fractional occupations of the
subsystems. 
We remark that this is a unique feature of the FDE theory, whereas
no such a dependence can be exploited in PDFT
and potential-functional embedding theory \cite{huang11} because in these cases the
occupation numbers are not defined a priori within the partitioning scheme
but are instead determined variationally.

Examining the variation of the total embedding energy 
with the subsystems' particle numbers we could provide
an explicit expression for the discontinuity of the
embedding potential at integer particles numbers. Thus, it was possible
to highlight the relevance of the kinetic derivative discontinuity 
in determining the behavior of the embedding potential,
which becomes evident as a cusp at integer particle numbers
for any approximate frozen-density embedding calculation.

Moreover, taking advantage of our definitions of
densities and chemical potentials at different subsystems' particle
numbers, we could introduce a rigorous definition of 
some important chemical descriptors within the FDE theory,
in agreement with previous work done in the context of PDFT\cite{cohen07}.
Thus, despite a full discussion of chemical reactivity theory was
well beyond the scope of this paper, we were able to indicate some 
important properties of the Fukui function and the global hardness
in relation to the embedding environment.

Finally, we remark that the present work constitutes an
important ground on which several future developments may be built.
In particular, we mention (1) the possibility to extend the present theory 
to the case of a generalized KS formalism,
so that also hybrid functionals can be considered; (2) the
consideration of total systems with ensemble densities;
and (3) the full investigation of the dependence of chemical
reactivity indexes on the environment.

\appendix
\section{Proof of Eq. (\ref{dedw})}
\label{appa}
In this appendix we provide a sketch of the proof of Eq. (\ref{dedw}).

Starting from Eqs. (\ref{aaa1}), (\ref{aaa2}), and (\ref{enadd}) and
using the chain rule for derivation we can write,
after a bit of algebra,
\begin{eqnarray}
\nonumber
&&\left(\frac{\partial E^{emb}}{\partial \omega}\right)^\pm =\\
\nonumber
&&= \int \left(\frac{\delta T_s[\rho_{A\omega}]}{\delta\rho_{A\omega}(\R_1)} + v_s[\rho_{A\omega}](\R_1) + v_{emb}^A(\R_1)\right)\left(\frac{\partial \rho_{A\omega}(\R_1)}{\partial \omega}\right)^\pm d\R_1 +\\
\nonumber
&& + \int \left(\frac{\delta T_s[\rho_{B\omega}]}{\delta\rho_{B\omega}(\R_1)} + v_s[\rho_{B\omega}](\R_1) + v_{emb}^B(\R_1)\right)\left(\frac{\partial \rho_{B\omega}(\R_1)}{\partial \omega}\right)^\pm d\R_1\ ,
\end{eqnarray}
where a $+$ in the superscript indicates that we are dealing with $\omega>0$,
whereas a $-$ denotes that we are considering $\omega<0$.
Because $\rho_{A\omega}$ and $\rho_{B\omega}$ are solutions
of the Euler equations defined by the first term in parenthesis
inside each integral, we can substitute these by the respective 
chemical potentials (see Eqs. (\ref{e17}) and (\ref{e18})). Hence,
\begin{equation}
\left(\frac{\partial E^{emb}}{\partial \omega}\right)^\pm = \mu_A^\pm\int
\left(\frac{\partial \rho_{A\omega}(\R_1)}{\partial \omega}\right)^\pm + \mu_B^\pm\int
\left(\frac{\partial \rho_{B\omega}(\R_1)}{\partial \omega}\right)^\pm\ .
\end{equation}

To complete the proof we note that 
\begin{eqnarray}
\nonumber
\int \left(\frac{\partial \rho_{A\omega}(\R_1)}{\partial \omega}\right)^\pm d\R_1 & = & \frac{\partial}{\partial \omega}\left( \int \rho_{A\omega}(\R_1)d\R_1\right)^\pm =\\
& = &  \frac{\partial (N_A+\omega)}{\partial \omega} = 1\\
\nonumber
\int \left(\frac{\partial \rho_{B\omega}(\R_1)}{\partial \omega}\right)^\pm d\R_1 & = & \frac{\partial}{\partial \omega}\left( \int \rho_{B\omega}(\R_1)d\R_1\right)^\pm = \\
& = & \frac{\partial (N_B-\omega)}{\partial \omega} = -1 \ .
\end{eqnarray}
Therefore, we finally find
\begin{equation}
\left(\frac{\partial E^{emb}}{\partial \omega}\right)^\pm = \mu_A^\pm-\mu_B^\pm \ . 
\end{equation}

\begin{acknowledgment}
This work was partially funded by the ERC Starting Grant FP7 Project DEDOM (No. 207441). We thank 
TURBOMOLE GmbH for providing the TURBOMOLE program package and M. Margarito for technical support.
\end{acknowledgment}


\end{document}